\documentclass[twocolumn,showpacs,preprintnumbers,amsmath,amssymb]{revtex4}
\usepackage{graphicx}
\usepackage{dcolumn}

\begin{document}
\title{Growth Kinetics Effects on Self-Assembled InAs/InP Quantum Dots}
\author{Bhavtosh Bansal\footnote{Email:bhavtosh@tifr.res.in}, M. R. Gokhale, Arnab 
Bhattacharya and B. M. Arora}
\affiliation{Department of Condensed Matter Physics and Materials Science, Tata Institute 
of Fundamental Research, 1 Homi Bhabha Road, Mumbai-400005, India.\\}
\date{\today}
\begin{abstract}
A systematic manipulation of the morphology and the optical emission properties of MOVPE 
grown ensembles of InAs/InP quantum dots is demonstrated by changing the growth kinetics 
parameters. Under non-equilibrium conditions of a comparatively higher growth rate and low 
growth temperature, the quantum dot density, their average size and hence the peak 
emission wavelength can be tuned by changing efficiency of the surface diffusion 
(determined by the growth temperature) relative to the growth flux. We further observe 
that the distribution of quantum dot heights, for samples grown under varying conditions, 
if normalized to the mean height, can be nearly collapsed onto a single Gaussian curve.      
\end{abstract}
\maketitle
Strained heteroepitaxy beyond the critical thickness can lead to spontaneous generation of 
three dimensional nanoclusters via the Stranski-Krastanov growth route \cite{stangl, 
sugawara}. The inherently statistical nature of this self-assembled growth process implies 
that quantum dots' areal density, average size and the dispersion in size around the 
average may be determined in subtle ways by the interplay of energetics and kinetics of 
the growth process. The morphological characteristics of the ensemble in turn determine 
the electronic density of states. Therefore understanding and experimentally controlling 
the size, density and the size dispersion of quantum dots has been a fundamental issue. 

In this letter, we have studied metal-organic vapor phase epitaxy (MOVPE) grown InAs/InP 
quantum dots. In a large number of previous studies on this system, the actual morphology 
and/or the emission properties of quantum dot ensembles have been found to be dependent on 
both the specific details of the material parameters and growth conditions (e. g., 
substrate miscut\cite{vicinal substrates}, long ranged surface stresses\cite{surface 
stress}, buffer morphology\cite{buffer layer morphology, buffer morphology}, the matrix 
material\cite{matrix material change1, matrix material change2}, material flux and partial 
pressures during growth\cite{3-5ratio}, annealing\cite{rta} and growth interruptions 
times\cite{growth interruption}) as well as on the generic growth parameters 
\cite{carlsson, marchand, ponchet_coverage} like the growth temperature, materials flux 
and coverage. Since the dependence of the quantum dots' properties on the latter set of 
growth kinetics parameters is expected to provide a more system independent insight into 
the problem of self-assembly, we have also followed this approach. Apart from the above 
mentioned references on the InAs/InP system, the approach also allows us to relate our 
work to other studies carried out in the same spirit but on different material systems. 
These include Monte Carlo simulations by Meixner, et al.\cite{meixner_kunert_scholl}, a 
rate equation based model for growth \cite{dobbs} and theoretical and experimental 
observations on InP/GaAs by Johansson and Seifert \cite{Johansson_Seifert1, 
Johansson_Seifert2} and many studies on the InAs/GaAs system, among which the one by 
Dubrovskii, et al.\cite{dubrovskii} is quite substantial. Furthermore, we demonstrate a 
very simple (phenomenological) scaling collapse of the {\em heights} distribution data 
onto a single Gaussian curve. The present study therefore attempts to demonstrate in a 
qualitative sense that the quantum dot density and the average size but also their size 
dispersion may be understood and therefore predicted on the basis of the three most basic 
growth parameters $-$coverage, growth rate and growth temperature$-$ provided the growth 
is carried out under far from equilibrium conditions.  

\begin{figure}[!th]
\begin{center}
\includegraphics[width=8.5cm,
  keepaspectratio,
  angle=0,
  origin=lB]{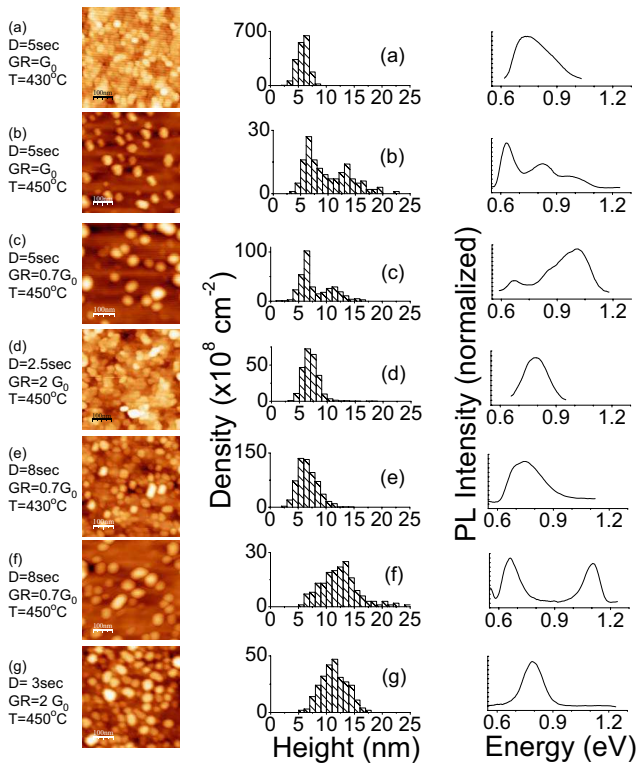}
\caption{\label{fig:fig1}{\it (left column)$\sim 500nm \times 500nm$ AFM surface scan 
images, (middle column) heights histograms and (right column) the 25K PL spectra (on the 
corresponding capped samples) grown at different growth rates (GR) and growth temperatures 
(T) and growth durations (D). Growth rate $G_0$ corresponds to approximately 2.5 ML/s. 
Peaks around 1eV in (b), (c) and (f) are due to wetting layer. Notice that the morphology 
and PL (a) and (d), and in (e) and (g) are qualitatively similar.}}
\end{center}
\end{figure} 

MOVPE growth was carried out on n+ doped (001) InP substrates using trimethyl indium and 
arsine as group III and V sources in a horizontal reactor at a pressure of 100 torr with 
hydrogen as the carrier gas. InAs layers were grown at a relatively low temperature of 
430-450$^\circ$C. Prior to InAs deposition, an InP buffer layer was grown, first $\sim$500 
\AA $\,$at 625$^\circ$C and then with temperature continuously ramped down to the InAs 
growth temperature and finally another 500 \AA$\,$ at the stable temperature. To avoid 
switching transients, the indium flux for the buffer, the InAs, and the cap layers was 
kept the same. For a given set of growth conditions, a pair of samples was grown with 
identically deposited InAs layer in two growth runs. In the first case, the sample was 
immediately cooled and taken out of the reactor after InAs deposition itself to enable a 
study of surface morphology and in the second case, an InP cap layer was grown for samples 
used for photoluminescence (PL) study. For these samples, about 50 \AA$\,$ InP cap was 
deposited at the InAs deposition temperature to minimise further ripening during the 
subsequent growth of the remainder of the cap at higher temperature. The uncapped dots 
were characterized by Nanoscope atomic force microscope in contact mode. PL spectra were 
recorded at $\sim$ 25K with a 0.67 meters McPherson grating monochromator and 325 nm 
helium-cadmium laser as the excitation source at a power density of $\sim$ 0.5W cm$^{-2}$. 
The measured spectra were corrected for the system response against a standard Oriel 
black-body source heated to 1350K.

Fig.\ref{fig:fig1} (a)-(g) summarizes the results of seven pairs of samples (uncapped for 
morphology and capped for PL measurements). The growth parameters (growth duration, 
approximate growth rate and growth temperature) for each case are given next to the 
figure. 
Six of the seven pairs of samples shown in the figure can be divided into two groups, 
where  the samples in Fig.\ref{fig:fig1}(a, b and d) and Fig.\ref{fig:fig1}(e, f and g) 
have nominally similar coverage but differing growth rates and growth temperatures. In the 
absence of in-situ thickness diagnostic tool for our MOVPE system, we have estimated the 
coverage as the trimethyl indium flux $\times$ growth duration, for a fixed group III 
source to group V ratio. This assumes that the growth rate scales linearly with the TMIn 
flux. The value of coverage was then inferred from the growth time. The growth rate was 
calibrated against the wetting layer PL peak\cite{wetting layer footnote} as described in 
a previous study\cite{our sk paper}. Since the primary confinement is along the growth 
direction, the peak and the dispersion in the heights histogram may be expected to 
proportionally show up in the low temperature (25K) PL spectrum of the corresponding 
capped sample, also depicted in Fig.\ref{fig:fig1} next to heights histogram. Despite the 
fact that overgrowth can substantially change the shape and composition of the dots and 
the areas sampled by PL and AFM are orders of magnitude different, there is a good 
qualitative agreement between the AFM data and the PL spectra. In particular, bimodally 
distributed dots show two PL peaks and the peak energy shifts with the average size of the 
dots. Possible reasons for a mismatch in the size dispersion with the PL linewidths are 
very briefly discussed later.

 From Fig. 1, it is evident that for similar coverage but by changing the growth rate and 
growth temperature, it is possible to change the quantum dots' density by over an order of 
magnitude. The corresponding peak PL emission wavelength is also seen to change from $\sim 
0.65 eV$ in Fig.\ref{fig:fig1}(f) to the more usual $\sim 0.8eV$ in Fig.\ref{fig:fig1}(a, 
d, e and g). At lower coverage, corresponding to intermediate stage of growth, we observe 
that the distribution is bimodal for comparatively smaller growth rates 
(Fig.\ref{fig:fig1}(b, c). This intermediate stage bimodality (Fig.\ref{fig:fig1}(b),(c)) 
is suppressed by making the growth more non-equilibrium, by either lowering the growth 
temperature as in Fig.\ref{fig:fig1}(a), or by increasing the growth flux as in 
Fig.\ref{fig:fig1}(d). This behavior has also been observed in previous studies on 
InAs/GaAs \cite{Saito} but also needs to be contrasted with the more complex trend 
observed for InP/GaAs samples \cite{Johansson_Seifert1}. 

Furthermore, we observe that Fig.\ref{fig:fig1}(a and d) and Fig.\ref{fig:fig1}(e and g) 
are qualitatively more similar to each other than they are to Fig.\ref{fig:fig1}(b) and 
Fig.\ref{fig:fig1}(f) respectively. This indicates that (1) a smaller growth rate (and an 
enhanced growth temperature) yields larger dots with a smaller areal density and (2) that 
the effect of a smaller growth rate can be compensated by a larger growth flux. 
Specifically,  our observation of point (2) is qualitatively very similar to the 
expectations in a recent growth simulation by Meixner, et al. (Fig. 7 in reference 
\cite{meixner_kunert_scholl}). The simplest models for self-assembled cluster 
growth\cite{dobbs, meixner_kunert_scholl} are developed in analogy with the submonolayer 
deposition\cite{barabasi} with the assumption that the later stage of self-assembly is 
largely dictated by the kinetic processes occurring at the surface. Then the average 
quantum dot density is dictated by how efficiently the preexisting material can diffuse 
and find an equilibrium site before more fresh material arrives on the surface. 
Quantitatively, this takes the form of a scaling relation\cite{barabasi, 
Johansson_Seifert2}, where the mean island density depends only on the ratio of the growth 
flux and surface diffusion efficiency. The largeness of this dimensionless ratio may also 
be taken to be the measure of departure from equilibrium.
\begin{figure}[!h]
\begin{center}
\resizebox{!}{7cm} {\includegraphics{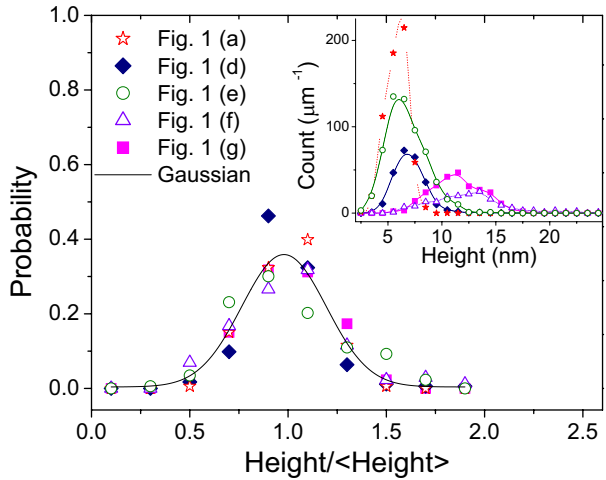}}
\caption{\label{fig:fig2}{\it Probability distribution function constructed from the 
respective histograms (Fig.\ref{fig:fig1}) scaled by the average height. Solid line shows 
a fit of the average of these data points to a Gaussian function. Inset shows the same 
histograms as in Fig.\ref{fig:fig1} except that the curve corresponding to Fig. \ref 
{fig:fig1} (a) is scaled down by a factor of three for clarity of comparison with other 
curves in the figure}}.
\end{center}
\end{figure} 

Although we have obtained quantum dots that, depending on the growth conditions, vastly 
vary in size and density, a striking feature of the height histograms in 
Fig.\ref{fig:fig1} (a, d, e, f and g) is that extent of size dispersion is proportional to 
the average height of the ensemble. Therefore we have re-plotted the heights histograms 
(for samples with a unimodal size distribution) in units of mean height\cite{collapse 
footnote} and normalized the area under the curve to unity in  Fig.\ref{fig:fig2}. Very 
approximately, we may describe the height dispersion by a single Gaussian curve centred 
around the mean ($h/ \langle h\rangle =0.98$) with a full width at half maximum of 0.43. 
These values provide a rough but very useful estimate of the expected size dispersion 
(since area $\propto$ height) in terms of the average size of dots. Since the average 
dot-sizes themselves may be written in terms of growth kinetic parameters, such a 
prescription can, {\em  in principle} lead toward a first principles prediction of the 
inhomogeneous broadening in terms of a few growth parameters, especially because the 
primary confinement occurs along the height of the quantum dots due to the large aspect 
ratios ($\sim 6-10$). 

While the peak energy and the modality of the size distribution can be correlated with the 
PL spectra, a direct correlation between the size dispersion and the low temperature PL 
emission linewidth is not always seen in the high density samples, Fig.\ref{fig:fig1}(a 
and e). This is presumably because of the strong interdot coupling effects. This may be 
understood as a combination of two effects (1) an overlap between the dots can lead to an 
excess `bandwidth' over and above the energy spread associated with quantum dots' size 
dispersion (2) despite a band formation (which would typically imply a narrower linewidth 
due to the transfer of carriers to the lowest available state in the density of states 
continuum), at low temperature, the strong potential fluctuations localize the excitons 
and they are not easily transferred to the lowest possible energy\cite{Runge}. A 
temperature dependent PL study that shows a qualitative difference between the temperature 
dependent emission properties of moderate and high density dots and which partially 
supports this hypothesis will be presented elsewhere. 

\noindent
{\em Conclusions:} MOVPE grown InAs/InP quantum dots ensembles grown at different growth 
rates and temperature were studied for their morphological and optical properties. We 
observed that the growth was largely kinetically determined with the surface diffusion 
being the most prominent process within the space of (deliberately highly non-equilibrium) 
growth parameters studied. The bimodality in quantum dots sizes and the PL emission peaks 
could be controlled by changing the growth conditions. It was also established, both in 
morphology and in optical properties, that the effect of lowering the growth temperature 
is qualitatively similar to the that of increasing the growth rate at a higher growth 
temperature. For dots with a unimodal distribution, the distribution of heights normalized 
by the average height were shown to be quite similar for samples with widely varying 
average heights (from 6nm to 12 nm). \\
\noindent
We thank J. John and Sandip Ghosh for their help with the AFM and PL measurements and 
Sandeep Krishna for his help with the development of the image processing software.  
\end{document}